\documentstyle[11pt,moriond,epsfig]{article}

\def\be{\begin{equation}}
\def\ee{\end{equation}}
\def\bea{\begin{eqnarray}}
\def\eea{\end{eqnarray}}

\newcommand{\xbj}{x} 

\newcommand{\SMALLXC}{SMALLXa,SMALLXb}
\newcommand{\CCFM}{CCFMa,CCFMb,CCFMc,CCFMd}
\begin{document}
\input feynman 
\bigphotons 
\noindent
     \hfill LUNFD6/(NFFL-7187) 2000 \\
\vspace*{3cm}
\title{PHOTON STRUCTURE AT HERA}

\author{H. JUNG }

\address{ Department of Physics, 
Lund University, 221 00 Lund, Sweden \\ 
\vspace{0.5cm}
For the H1 and ZEUS Collaborations
}

\maketitle\abstracts{The structure of the virtual photon and its contribution to
small $\xbj$ processes in deep inelastic scattering at HERA is discussed.}

\section{Introduction}
In electron-proton scattering the internal structure of the proton as  
well as of the exchanged photon can be resolved provided the scale $\mu^2$ of  
the hard subprocess is larger than the inverse size of the proton,  
$1/R^2_p \sim \Lambda_{QCD}^2$, and inverse size of the photon,
 $1/R^2_{\gamma} \sim Q^2$, 
respectively. Resolved photon processes play an important role in  
photo-production of high $E_T$ jets \cite{H1_resgamma_00,ZEUS_resgamma_99},
 where $Q^2 \approx 0$,  
but they can also give considerable contributions to DIS  
processes \cite{H1_incl_jets,Chyla_res_gamma}.  
if the scale $\mu^2$ of the hard  
subprocess is larger than $Q^2$. 
In the following I shall discuss the role of resolved virtual photons in small
$\xbj$ $ep$ scattering.

\section{Small $\xbj$ parton dynamics and the structure of the virtual photon}
The cross section 
at low $\xbj$ and large $Q^2$ with a high $p^2_T$ jet in the  
proton direction (a forward jet)  has been 
advocated 
as a particularly sensitive measure of small $\xbj$ parton dynamics 
\cite{Mueller_fjets1,Mueller_fjets2}. If the forward
jet has large energy ($x_{jet}=E_{jet}/E_{proton} \gg \xbj$) 
the evolution from $x_{jet}$ to small $\xbj$ can be studied.
When $E_T^2 \sim Q^2$ there is no room for $Q^2$ evolution left and the DGLAP
formalism predicts a rather small cross section in contrast to the BFKL
formalism, which describes the evolution in $\xbj$. More information about the
underlying parton dynamics can be obtained by a measurement of the forward jet
cross section 
 as a function of the
ratio $E^2_T/Q^2$.
Three interesting regions of phase space can be defined:
\begin{itemize}
\item $E^2_T < Q^2$ ``standard" DIS region (direct photon processes), $Q^2$ is
the largest scale in the process
\item $E^2_T \simeq Q^2$ BFKL region, two scale problem.
\item $E^2_T > Q^2$ ``resolved $\gamma^*$" region, $E^2_T$ is the largest scale,
similar to photo-production processes.
\end{itemize}
\begin{figure}[htb]
\vskip -1cm
\hbox{
\vbox{\psfig{figure=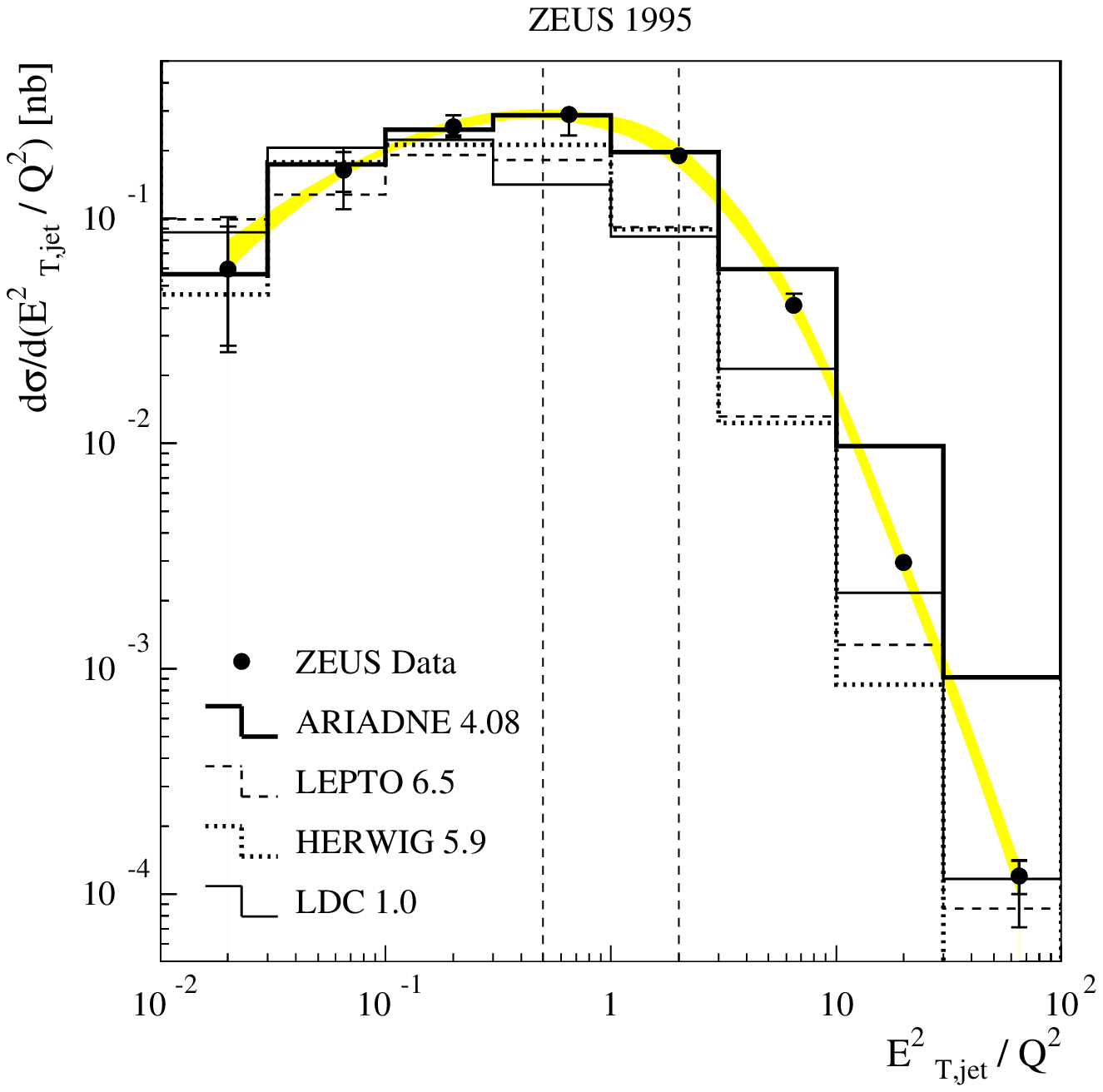,width=8cm,height=8cm}}
\hspace*{-8.0cm}
\vbox{\psfig{figure=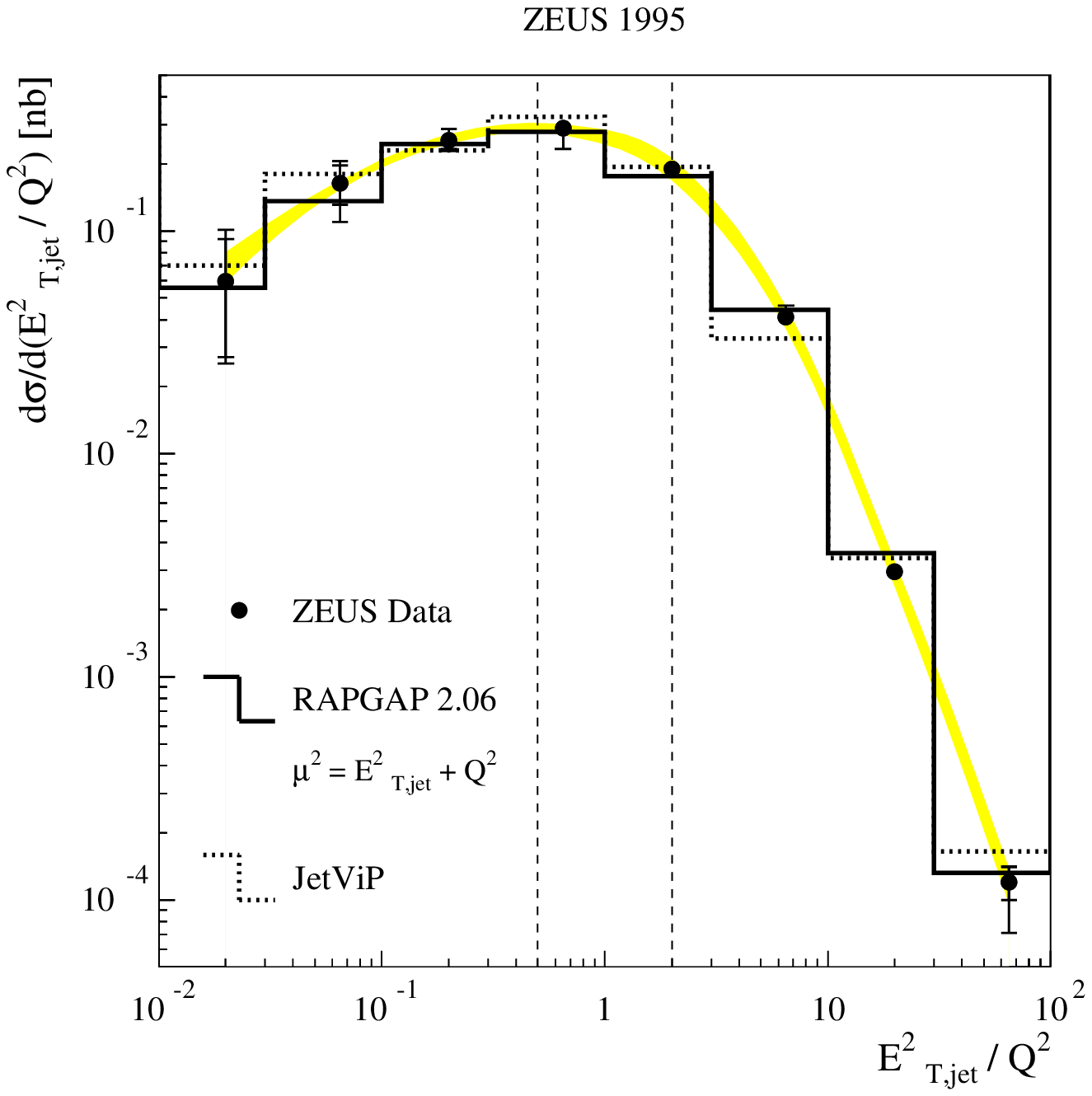,width=8cm,height=8cm}}}
\vskip -7.0cm 
\hspace*{6cm}(a)
\hspace*{8cm}(b)
\vskip 1.5cm
\hspace*{1.3cm} {\tiny ``direct $\gamma^*$" }
\hspace*{0.75cm} {\tiny BFKL}
\hspace*{0.9cm} {\tiny ``resolved $\gamma^*$" }
\hspace*{2.1cm} {\tiny ``direct $\gamma^*$" }
\hspace*{0.75cm} {\tiny BFKL}
\hspace*{0.9cm} {\tiny ``resolved $\gamma^*$" }
\vskip 4.0cm 
\caption{The forward jet cross section as a
function of $E^2_T/Q^2$ in the region $Q^2>10$ GeV$^2$, $\eta_{jet} < 2.6$,
$E_{t\;jet} > 5$ GeV, $x_{jet}>0.036$. The shaded band shows the hadronic energy
scale uncertainty. In $a$ predictions from standard deep inelastic Monte Carlo
generators are shown. In $b$ the prediction from a Monte Carlo generator
and a full NLO calculation including resolved virtual photons are shown
\protect\cite{fwdjet_potter}.
 Indicated are also the three different regions of
phase space defined in the text.
\label{fwd_jet_1}}
\end{figure}
The forward jet cross section as a function of $E^2_T/Q^2$ 
\cite{ZEUS_fjets_pt2/q2} is shown in Fig.\ref{fwd_jet_1}. In
Fig~\ref{fwd_jet_1}$a$ the data are compared with predictions from 
standard deep inelastic Monte Carlo programs, LEPTO~\cite{Ingelman_LEPTO65} and 
HERWIG~\cite{HERWIG} which use direct
photon processes only. The cross section in the
standard deep inelastic region $E^2_T < Q^2$ is
reasonably well described, but in the other regions underestimated by large
factors.
  In
Fig~\ref{fwd_jet_1}$b$ the data are compared with predictions including also
 processes from
resolved virtual photons. The data are well described over
the full phase space. 
It should be noted that only the anomalous component of the resolved virtual
photon contributes significantly at large $Q^2$
(which can be calculated in pQCD), whereas the
vector meson component becomes negligible.
Thus a reasonable approximation to small $\xbj$ parton dynamics can be achieved
by adding a resolved virtual photon contribution to the direct photon processes.
By doing this, a situation similar to BFKL is obtained:
The transverse momenta of the emitted partons need no longer be ordered in
transverse momentum, since in a resolved photon process there are two DGLAP
ladders, one from the photon and one from the proton towards the hard scattering
matrix element.
 However in order to achieve such a good description of the
data, the scale $\mu^2$ for the hard subprocess has to be sufficiently large,
e.g. 
$\mu^2 = Q^2 + p_T^2$
or $\mu^2 = 4 p_T^2$, as was shown in~\cite{jung_resgamma}. 
If the scale is chosen
to be $\mu^2 = p_T^2$, the predictions fall below the data. In that case there
is only a small contribution from resolved virtual photons. 
Thus to successfully also
describe the BFKL region, a (artificial) sufficiently large
scale $\mu^2$ has to be chosen to simulate enough high $p_T$ emissions.
\section{Structure of virtual photons}
\begin{figure}
\vskip -1cm
\hspace*{-0.5cm}
\hbox{
\vbox{\psfig{figure=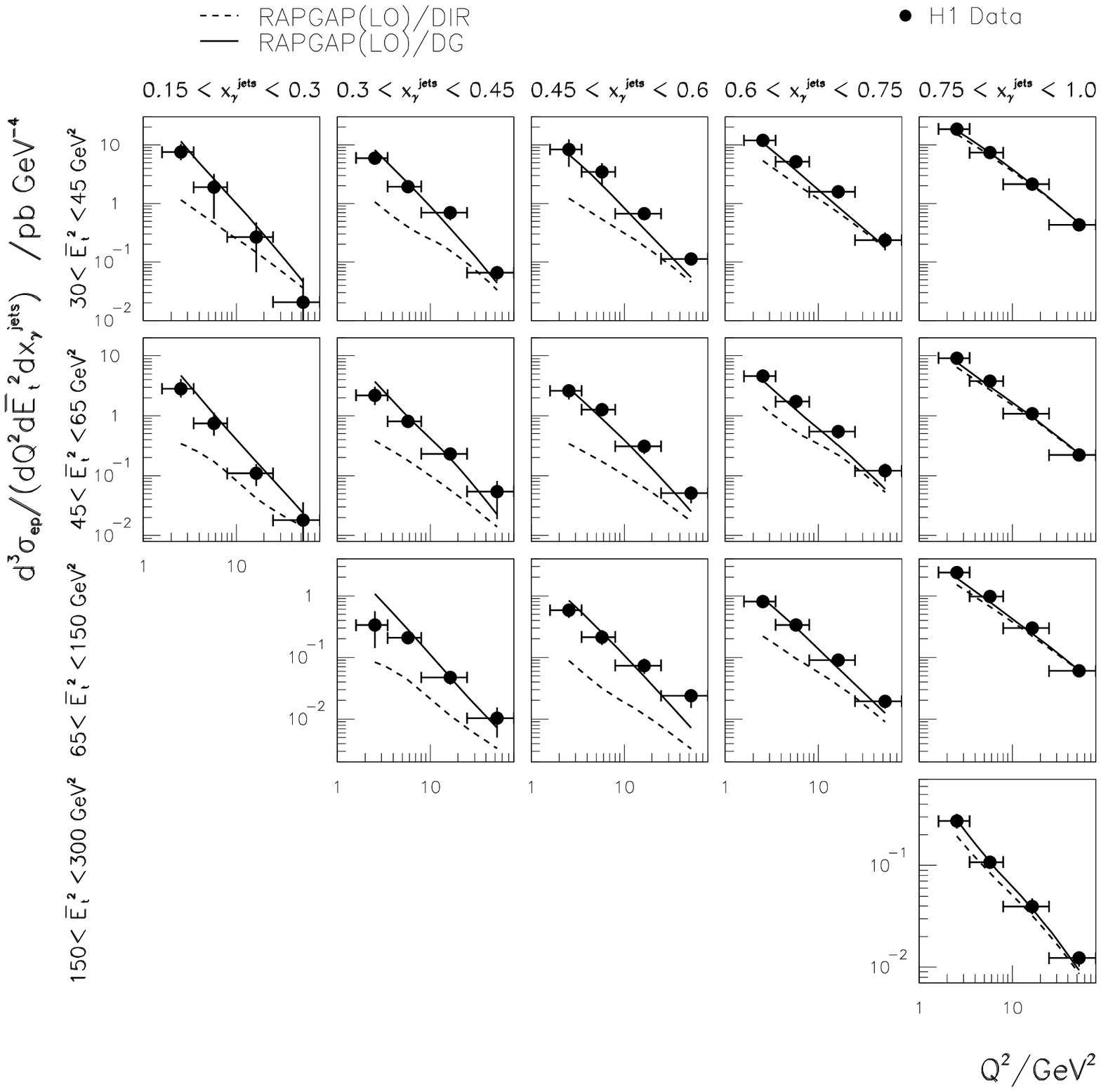,height=8cm,width=8cm}}
\hspace*{-8.0cm}
\vbox{\psfig{figure=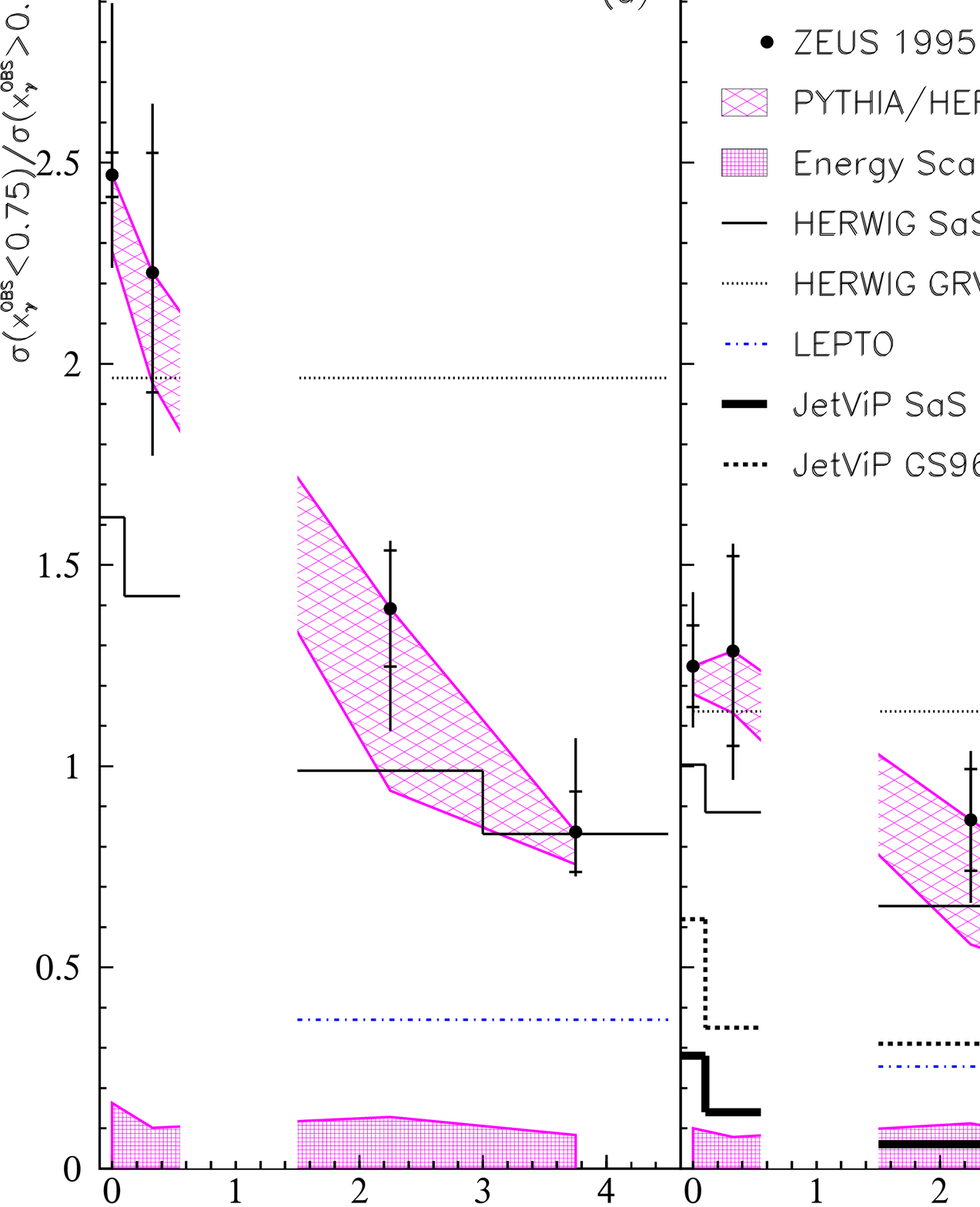,height=7.5cm,width=8cm}}}
\vskip -0.0cm 
\hspace*{6cm}(a)
\hspace*{8cm}(b)
\vskip 0.0cm
\caption{$a.$ 
The cross section of di-jet production~\protect\cite{H1_resgamma_dis}
 as a function of $Q^2$ in bins
of $x_{\gamma}$ and $E_T^2$ in the kinematic range of $0.1 < y < 0.7$ and
di-jets found with the inclusive $k_T$  jet algorithm with 
$\bar{E}_T^2 > 30$~GeV$^2$. Also shown is the Monte Carlo prediction for the
direct photon contribution alone (dashed line) and the sum of direct and
resolve photon processes (solid line).
$b.$ The ratio 
$R=\sigma^{resolved}/\sigma^{direct}$~\protect\cite{ZEUS_resgamma_dis}
 as a function of $Q^2$
for the low and high $E_T$ sample. Also shown are the predictions
 from the HERWIG
Monte Carlo program including a contribution of resolved photons
and a NLO calculation~\protect\cite{jetvip}.}
\label{H1_resgamma_dis}
\end{figure}

In photo-production the high $E_T$ di-jet cross section has been 
extensively used
to study the structure of the real photon as a function of $x_{\gamma}$,
in LO the fraction of the photon momentum carried by the 
parton~\cite{H1_resgamma_00,ZEUS_resgamma_99}. High
$E_T$ di-jet production at $Q^2> 0$ can be used to determine the structure of
the virtual photon, if $E_T^2 > Q^2$. Such studies have been performed by
H1~\cite{H1_resgamma_dis} and ZEUS~\cite{ZEUS_resgamma_dis}. In
Fig.~\ref{H1_resgamma_dis}$a$
 the di-jet cross section is shown as a function of
$Q^2$ for different bins in $x_{\gamma}$ and $E_T$ together with a Monte Carlo
prediction of direct photon processes and the sum of direct and resolved virtual
photons. In the region of small $x_{\gamma}$ and small $Q^2$ the measured cross
section lies far above the prediction involving direct virtual photons only.
This indicates that additional processes are needed to explain the data. 
Including in addition contributions from resolved virtual
photon processes a reasonable description of the data is obtained.
 The contribution
of resolved virtual photons is largest for $Q^2 \ll E_T^2$ and
$x_{\gamma} \ll 1$.
A complementary measurement has been performed by ZEUS~\cite{ZEUS_resgamma_dis}
 measuring di-jet
cross section in the region $0 < Q^2 < 4.5$~GeV$^2$ for two different regions
in $x_{\gamma}$, the direct photon region ($x_{\gamma}>0.75$) and the resolved
photon region  ($x_{\gamma}<0.75$). In Fig.~\ref{H1_resgamma_dis}$b$
 the ratio
$R=\sigma^{resolved}/\sigma^{direct}$ as a function of $Q^2$ is shown for a low
$E_T$ ($E_T > 5.5$ GeV) and a high $E_T$ ($E_{T\,1} >7.5$ GeV and 
$E_{T\,2} >6.5$ GeV) sample.
The high $E_T$ sample is much less influenced by multiple interactions compared
to the low $E_T$ sample.
Also here a significant contribution over
the standard direct photon processes are needed to describe the data.
Thus we can conclude that in the region of $E_T^2 > Q^2 > 0$ direct photon
processes are not sufficient to describe the data. Additional
processes, like the contribution of resolved virtual photons, are needed to
improve the description of the data.
\section{What did we learn?}
We have seen that the concept of resolved virtual photons provides a good
description of di-jet production in the region where $E_T^2 > Q^2$. However in
general in
DIS the situation is more complicated than in photo-production, because of the
presence of two hard scales: $E_T^2$ and $Q^2$.
DIS processes may be separated into 
three different regions of phase space, depending whether $Q^2$ or $p_T^2$ is
larger or both are of the same order. From the discussion of the forward jet
cross section as function of $E_T^2/Q^2$ it became clear,
 that the full phase
space can be described using the concept of resolved virtual photons provided
 the scale of the hard subprocess is large enough: $\mu^2 > E_T^2$. 
\par
Thus, if we consider $\mu^2 = E_T^2$ to be the proper hard scale, then
 deep inelastic scattering turns out to consist of three different pieces:
the standard DIS region (pointlike photon) with $Q^2$ being the largest scale, a
typical resolved photon region where $E_T^2$ is the largest scale, and then the
typical BFKL region where $Q^2 \sim E_T^2$. Such a picture 
is quite unsatisfactory
since it requires a proper treatment of the interplay of the contributions in
the different regions of phase space as well as the problem of possible 
double counting. However all the difficulties in describing the data come from
the approximations used in the evolution of the parton densities and the
calculation of the hard scattering matrix elements: the leading $\log Q^2$ -
or collinear approximation. These problems can be overcome by using the $k_T$ -
factorization or semi-hard approach together with the CCFM~\cite{\CCFM}
 evolution equations
for the parton densities, which reproduces the DGLAP and the BFKL equation
in the appropriate limits. Since in the $k_T$ - factorization approach the
matrix elements are treated with off shell incoming partons, the anomalous
component of the resolved photon is automatically included as soon as $k_T^2$
becomes larger than $Q^2$. 
Recently it has been shown~\cite{\SMALLXC,CASCADE} that 
the cross section of
forward jet production as a function of $E_T^2/Q^2$ can be nicely reproduced
within the CCFM framework. This gives confidence that we are on the way to
obtain a unified picture of high energy deep
inelastic lepton scattering.
\section*{Acknowledgments}
I am grateful to the organizers for this stimulating conference. I am also
grateful to the Swedish Science Council (NFR) for financial support.

\end{document}